\documentclass[reprint,aps,pra,superscriptaddress]{revtex4-1}

\usepackage{textcomp}
\usepackage{amsmath}
\usepackage{amssymb}
\usepackage{mathrsfs}
\usepackage[normalem]{ulem}
\usepackage{color}
\usepackage{empheq}
 
\usepackage{graphicx}

\begin{document}

\title{Benchmarking quantum annealers using symmetries in embedded subgraphs}

\author{Dilina Perera}
\affiliation{Department of Physics and Astronomy, Mississippi State University, Mississippi State, Mississippi 39762-5167, USA}
\affiliation{Department of Physics, University of Colombo, Colombo 03, Sri Lanka}

\author{Bhavika Bhalgamiya}
\affiliation{Department of Physics and Astronomy, Mississippi State University, Mississippi State, Mississippi 39762-5167, USA}

\author{M. A. Novotny}
\affiliation{Department of Physics and Astronomy, Mississippi State University, Mississippi State, Mississippi 39762-5167, USA}
\affiliation{HPC\textsuperscript{2} Center for Computational Sciences, Mississippi State University, Mississippi State, MS 39762-9627, USA}
\affiliation{Faculty of Mathematics and Physics, Charles University in Prague, Ke Karlovu 5, CZ-121 16 Praha 2, Czech Republic}

\begin{abstract}

We investigate an efficient, generic method for evaluating the performance of quantum annealing devices
that does not require the prior knowledge of the true ground state of the benchmark problem.
This approach exploits symmetry properties inherent to the ground states of a composite Hamiltonian comprising 
the benchmark problem Hamiltonian and its symmetric counterpart.
Using this method, we compare the performance of two generations of D-Wave machines. 
Although we do not observe a noticeable difference in the probability of finding solutions with the required symmetry,
our results suggest that the current generation of D-Wave machines notably outperforms its predecessor 
in finding states closer to those with the required symmetry.

\end{abstract}


\maketitle

\section{Introduction}
Optimization is a continuously evolving branch of mathematics and computer science that has applications in diverse fields such as
industrial engineering, materials sciences, finance, machine learning etc.
The common goal is to determine the optimal solution that minimizes a given cost function.
Due to many local minima separated by high barriers in the cost function landscape,
the worst-case instances of archetypical optimization problems
such as the traveling salesman problem, Boolean satisfiability ($k$-SAT) problem, knapsack problem etc.
cannot be solved in polynomial time,
unless the polynomial and nondeterministic polynomial complexity classes are identical ($\text{P} = \text{NP}$).
These problems share common ground with an age-old problem in statistical physics, 
namely, finding the ground state configuration of an Ising spin glass~\cite{binder:86, nishimori:01} with frustrating interactions.
Consequently, many of the modern heuristic techniques for optimization have been inspired by concepts and methods in classical statistical physics; (thermal) simulated annealing (SA)~\cite{kirkpatrick:83}, 
parallel tempering~\cite{swendsen:86, hukushima:96, katzgraber:01, katzgraber:06}, 
and population annealing~\cite{hukushima:03, machta:10, wang:15, wang:15a} to name a few.
In the widely used technique of SA~\cite{kirkpatrick:83}, 
one introduces thermal fluctuations with the aid of a real or fictitious temperature variable,
allowing the system to hop over energy barriers.  
The temperature is slowly reduced to a target value close to zero, with the expectation that the system would gradually settle down
to the global energy minimum. 

An alternative annealing scheme is based on quantum mechanical principles, using quantum fluctuations and quantum properties to find optimal solutions, 
and is called quantum annealing (QA)~\cite{finnila:94, kadowaki:98, santoro:02}. 
The quantum approach has recently been implemented in hardware~\cite{johnson:11, barends:16}, and further rapid advances in hardware for QA is expected.  
QA has been shown to be equivalent to gated quantum computation~\cite{mizel:07, biamonte:08, aharonov:07}.  
For additional references and a history of QA, see recent reviews~\cite{albash:18, das:08}. 

Hardware-realized QA is still an emerging technology, and the only hardware vendor is D-Wave Systems Inc.  
D-Wave provides programmable machines, which strive to solve optimization problems in 
the quadratic unconstrained binary optimization (QUBO) form.  
Consequently, in this paper we present our benchmarking analysis of two generations of D-Wave machines, 
the 1000-qubit previous generation machine (model 2X) and the 2000-qubit current generation machine (model 2000Q).  
The native graph on these machines is the $K_{4,4}$ Chimera graph, imposing 
a restrictive connectivity (See FIG.~\ref{fig:chimera_lattice}).  
These devices suffer from a number of limitations, such as not having a finite-temperature 
spin-glass transition due to the native graph~\cite{katzgraber:14}, 
and perturbations in the problem Hamiltonian  due to noise and engineering constraints~\cite{zhu:16, pudenz:14}.  
Although it remains an open question as to whether the device has any definitive ``quantum advantage'' 
over state-of-the-art classical algorithms, 
the device has demonstrated a limited speedup over selected classical solvers for certain specifically-designed 
synthetic problems~\cite{mandra:18, albash:18b}.
Moreover, comparisons with quantum and classical models suggest that the device 
does exhibit quantum behavior~\cite{albash:15, boixo:14, wang:13, boixo:13}.
A study of small-scale systems of superconducting flux qubits has also found
experimental evidence for quantum entanglement during the annealing process~\cite{lanting:14}.

An abundance of studies have been devoted to benchmarking D-Wave devices against 
conventional optimization schemes~\cite{ronnow:14, boixo:14, king:15, hen:15, king:15b, venturelli:15, mandra:16, denchev:16, king:19, mandra:18, albash:18b}.
The common approach for benchmarking is to compare optimal median time to solution~\cite{ronnow:14, boixo:14} 
determined using the success probability of obtaining ground states for a problem ensemble.
This requires problem instances with pre-determined ground states.
Apart from classes of problems for which the exact ground states are \textit{a priori} 
known (i.e., planted solutions)~\cite{hen:15, king:15b, hamze:18, hen:19, hamze:19}, 
the ground states of the benchmark problem instances are generally determined 
using highly-optimized implementations of classical optimization algorithms~\cite{wang:15a, katzgraber:15}.
This process requires a significant amount of computational resources and time.
Moreover, because of the heuristic nature of the optimization schemes, there is no guarantee that 
the obtained solutions are indeed the true ground states of the problem Hamiltonian.

In this paper, we demonstrate an alternative way of evaluating the performance of QA devices
based on spatial symmetries in embedded graphs.
In a recent preliminary study~\cite{perera:16}, we briefly explored the applicability of this method
for checking the validity of candidate ground state solutions returned by QA devices.
The method does not require the true ground states of the problem instances to be known in advance,
and hence presents a significant advantage over the conventional benchmarking scheme with regard to computational overhead.
Moreover, the method does not depend on the details of the device's native architecture or the underlying technology.
Thus it can be used with future generations of QA devices with arbitrary native graph or hypergraph
structures, as well with programmable Ising solvers based on alternative technologies,
for example, coherent Ising machines~\cite{wang:13b,hamerly:18x} 
and the Fujitsu Digital Annealer~\cite{matsubara:17,tsukamoto:17}. 

\section{The method}

\begin{figure}[h] 
  \includegraphics[width=0.8\linewidth]{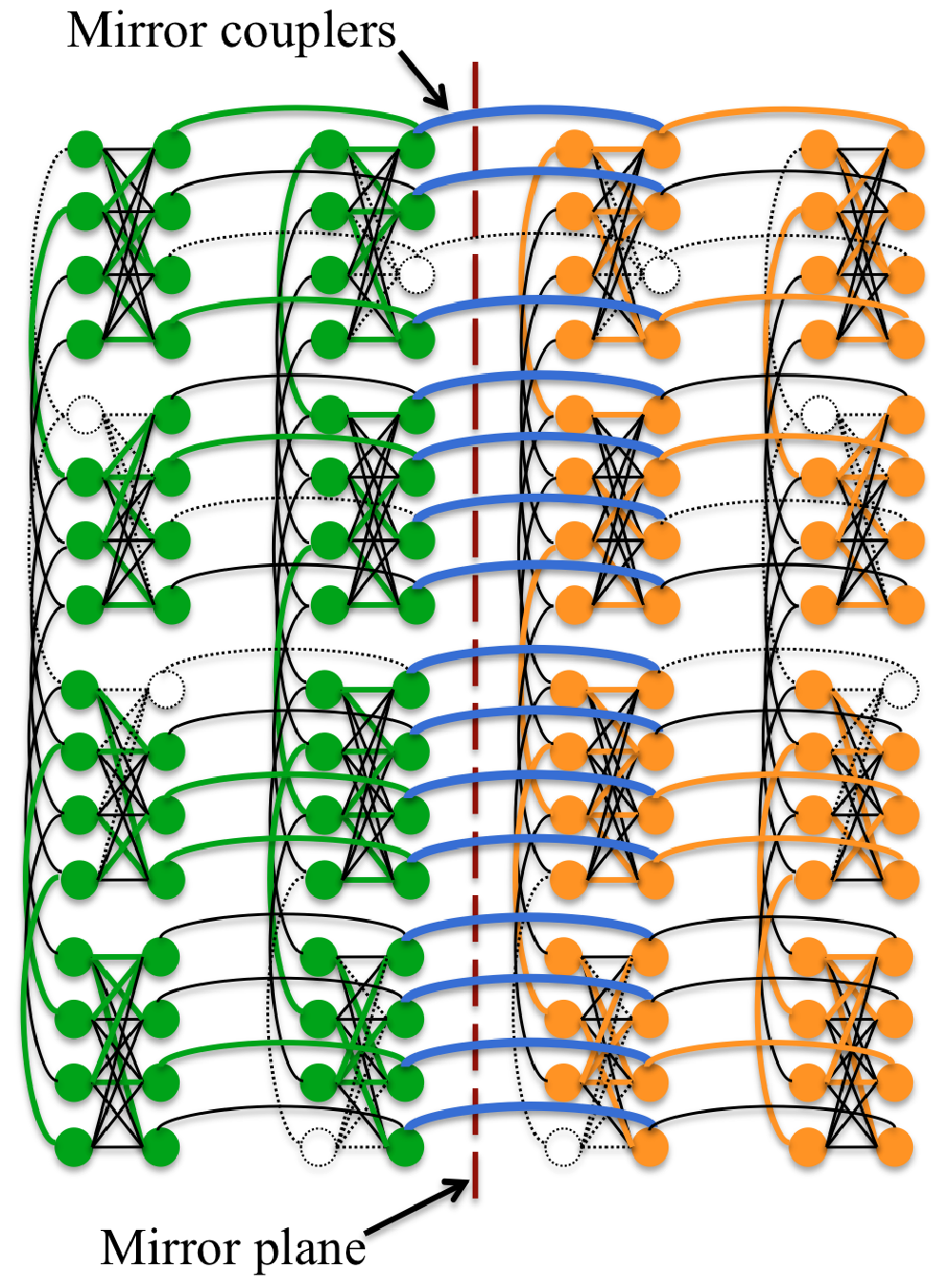}
  \caption{Application of the method on a quantum annealing device with a native Chimera topology,
		with a $4 \times 2$ Chimera graph as the problem graph $G$.
		Graph $G$ (green) and its mirror image $G^\prime$ (orange) are, respectively, embedded on the left and right sides of the mirror plane.
		The horizontal couplings that connect the $2$nd and the $3$rd columns of Chimera unit cells are designated as mirror couplings (blue).
	 	White-colored qubits and couplings marked as dotted lines represent the device's inaccessible qubits and couplings and their mirror counterparts.
	 	}		
 \label{fig:chimera_lattice}
\end{figure}

The foundation of our benchmarking method is to simultaneously evolve two subgraphs based on spatial or spatial+spin-inversion symmetry of the QA device's native graph, 
when the graph is embedded onto a Euclidean space.  
The technique can be extended to more than two subgraphs evolved simultaneously, and to other symmetries.  
Moreover, the method can also be applied to hypergraphs for the case of Hamiltonians with many-body interactions.
In this paper we concentrate on two subgraphs, and on mirror or mirror+spin-reversal symmetry relating the two embedded subgraphs.  
As in FIG.~\ref{fig:chimera_lattice}, the two subgraphs are the mirror image of each other,
and a selected set of qubits in one subgraph is either ferromagnetically or antiferromagnetically coupled to the corresponding mirror qubits in the second subgraph
in order to enforce symmetry constraints on the ground state solutions of the composite Hamiltonian.
Candidate solutions that do not satisfy these symmetry constraints can then be eliminated as invalid.
To demonstrate this process, let us consider a QUBO problem that has binary variables 0 and 1.
In this paper, we describe these problems using Ising variables,
with the resultant Ising spin-glass Hamiltonian given by 
\begin{equation} \label{eq:classical_hamiltonian}
	\mathcal{H}_\text{prob} = -\sum_{(i,j) \in E} J_{ij} S_i S_j - \sum_{i \in V} h_i S_i \qquad S_i \in \{\pm 1\},
\end{equation}
where the subgraph $G = (V,E)$ with vertices $V$ and edges $E$ captures the structural information pertaining to the problem,
with $J_{ij}$ and $h_i$ being the exchange couplings and the local fields, respectively.
Let $\mathcal{H}_\text{prob}^\prime$ be a copy of $\mathcal{H}_\text{prob}$ with subgraph $G^\prime = (V^\prime,E^\prime)$,
which is a mirror image of $G$ with respect to the chosen mirror plane.
A subset of vertices $\{k\} \subset V$ in $G$ are directly connected to their mirror counterparts 
$\{k^\prime\} \subset V^\prime$ in $G^\prime$ 
using ``mirror couplings'', all with the same sign and of strength $M_k$, as represented by the coupling Hamiltonian
\begin{equation}
	\mathcal{H}_M = - \sum_k M_k S_k S_{k^\prime}.
\end{equation}
The resulting final Hamiltonian takes the form
\begin{equation} \label{eq:classical_hamiltonian}
	\mathcal{H}_T = \mathcal{H}_\text{prob} + \mathcal{H}_\text{prob}^\prime + \mathcal{H}_M.
\end{equation}

FIG.~\ref{fig:chimera_lattice} illustrates the application of this method on
a QA device with a native Chimera topology, 
where we have chosen a $4 \times 2$ $K_{4, 4}$ Chimera graph as the original problem graph $G$.

Let $\left(q_k, q_k^\prime\right)$ represent a pair of physical qubits connected with a mirror coupling.
In the case of ferromagnetic mirror couplings ($M_k > 0$) with sufficiently large magnitude,
the coupling term $\mathcal{H}_M$ imposes the constraint $q_k = q_k^\prime$ on all $\left(q_k, q_k^\prime\right)$ pairs during the annealing process.
In the case of antiferromagnetic mirror couplings ($M_k < 0$), the constraint $q_i = -q_i^\prime$ is imposed,
and the spin reversal symmetry requires the signs of all local fields on the two subgraphs to be different.
These constraints, in turn, impose symmetry requirements on the ground state solutions of the composite Hamiltonian $\mathcal{H}_T$.
This can be easily understood by considering the simplest scenario in which the ground state of the problem Hamiltonian $\mathcal{H}_\text{prob}$ is non-degenerate.
If the solution returned by the device is indeed the true ground state of the composite Hamiltonian $\mathcal{H}_T$,
for $M_k > 0$, the corresponding classical spin configuration will have reflection symmetry with respect to the mirror plane.
For $M_k < 0$, the spin configuration will have spin-flip (``up-down'') symmetry with respect to the mirror plane.

If the ground state of $\mathcal{H}_\text{prob}$ is degenerate, $\mathcal{H}_T$ will also have degenerate ground states,
and not all of them may satisfy the aforementioned symmetry constraints.
This is due to the fact that it is possible to have ground states of $\mathcal{H}_T$ comprised of two different ground state configurations of 
$\mathcal{H}_\text{prob}$ and $\mathcal{H}_\text{prob}^\prime$ respectively occupying the two sides of the mirror plane, 
which also happen to satisfy the constraints imposed by $\mathcal{H}_M$ on the coupled qubit pairs $\left(q_k, q_k^\prime\right)$.
This scenario may not be uncommon for problems with highly degenerate ground states, 
such as for which the exchange couplings are drawn from the bimodal distribution, i.e. $J_{ij} \in \{\pm 1\}$.
However, we claim that the probability for such asymmetric ground states to result from the annealing process is significantly low,
since the spatial correlations induced by $\mathcal{H}_M$ increase the likelihood of symmetrical states across the mirror plane.

Note the presence of reflection/spin-flip symmetry in the solution does not necessarily imply 
that the ground state of $\mathcal{H}_T$ has been realized.
That is, occasionally, an excited state of $\mathcal{H}_\text{prob}$ and its mirror counterpart may occupy the two sides of the mirror plane, 
satisfying the same symmetry conditions.
Hence, the presence of symmetry should only be regarded as a measure that increases one's expectation that the true ground state has been realized,
rather than definite proof of such a realization.

We point out that our method bears some resemblance with a recently introduced quantum annealing error correction scheme~\cite{pudenz:14, pundez:15}
with regard to simultaneously evolving multiple, coupled copies of the problem Hamiltonian.
However, our method significantly differs from the error correction scheme 
in that we impose additional physical constraints on the graph embedding process 
to make use of the spatial symmetries.

\section{Results}

We now apply the answer checking method to evaluate the performance of four quantum annealing devices:
the previous-generation, 1000-qubit D-Wave 2X device (DWP) and three versions of the current-generation, 
2000-qubit D-Wave 2000Q devices, namely, an early experimental device hosted at D-Wave Systems (DWC),
the device currently hosted at NASA Ames Research Center (DWC [NASA]),
and a new lower-noise device hosted at D-Wave Systems (DWC [lower noise]).
The native Chimera graph of each device was split into two subgraphs of equal size via a horizontal mirror plane.
For the current-generation $16 \times 16$ D-Wave 2000Q devices, the dimensions (in units of 8-qubit unit cells) 
of the largest subgraphs are $16 \times 8$, whereas for the $12 \times 12$ DWP device, the largest graph size is $12 \times 6$.
To make the two subgraphs identical, inaccessible qubits and couplings on each side of the mirror plane 
were mirrored onto the other side.
As the problem graphs, we used Chimera graphs with different sizes,
with the values of the exchange couplings randomly drawn from the Sidon set $S_{28}$~\cite{zhu:16, katzgraber:15}, 
i.e. $J_{ij} \in \{\pm 8/28, \pm 13/28, \pm 19/28, \pm 1\}$.
The local fields were either set to zero or randomly drawn from the same Sidon set.
The problem graphs and their mirror counterparts are embedded into the respective subgraphs 
such that they are adjacent to the mirror plane. 
The horizontal couplings that span across the mirror plane were designated as mirror couplings (see FIG.~\ref{fig:chimera_lattice}),
and their values were set to the maximum ferromagnetic value of $+1$ allowed on the D-Wave machines.
We also repeated some points with antiferromagnetic values, all $M_k=-1$, with results within the errors obtained for the $M_k=+1$ results.
For each problem graph size, $1000$ random instances with different coupling values were generated.
For each instance, we performed $1000$ annealing runs, and examined the solution/solutions that correspond to the lowest energy.
If at least one of the lowest-energy solutions was found to have reflection symmetry about the mirror plane,
we speculated that the true ground state may had been achieved.
Based on the results of the $1000$ random instances, we obtained an estimate of the probability $P_\text{sym}$ that at least one of the lowest-energy solutions 
was found to have reflection symmetry.

\begin{figure}[h] 
  \includegraphics[width=\linewidth]{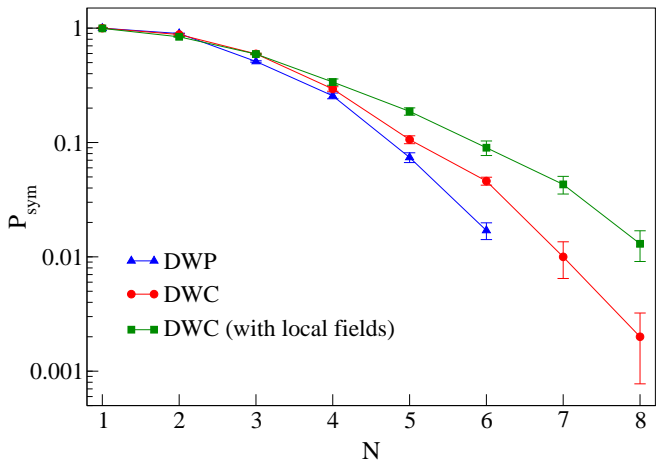}
  \caption{
		Probability ($P_\text{sym}$) that out of $1000$ annealing runs, at least one of the lowest-energy solutions 
		is found to satisfy symmetry requirements imposed by the mirror couplings. 
		Note the logarithmic scale for $P_\text{sym}$.
		The number of rows of unit cells in the problem graphs was fixed to $12$, 
		while the number of columns $N$ was varied from $1$ to $6$ for the DWP, and from $1$ to $8$ for the DWC
		and DWC [NASA].
		The blue, red, and brown curves, respectively, show the results obtained for the DWP, DWC, and DWC [NASA] for graphs
		with zero local fiends and couplings drawn from the Sidon set $S_{28}$.
		The green and purple curves, respectively, show the results obtained for the DWC and DWC [NASA] 
		for graphs with both couplings and local fields drawn from the Sidon set.
		For all calculations, the default annealing settings of the devices were used.
	  }		
 \label{fig:success_prob_1}
\end{figure}

FIG.~\ref{fig:success_prob_1} shows $P_\text{sym}$ for problem graphs with different sizes on the DWP, DWC, and DWC [NASA].
The number of rows of unit cells was fixed to $12$ for all graphs, 
while the number of columns $N$ was varied from $1$ to $6$ for the DWP, and from $1$ to $8$ for the DWC and DWC [NASA].
The blue, red, and brown curves, respectively, compare $P_\text{sym}$ for the DWP, DWC, and DWC [NASA]
for graphs with zero local fields and the exchange couplings randomly drawn from the Sidon set.
The green and purple curves show $P_\text{sym}$ for the DWC and DWC [NASA] for graphs with both couplings 
and local fields drawn from the Sidon set.
For all curves, $P_\text{sym}$ rapidly decreases with increasing $N$,
which is consistent with the fact that the number of qubits in the problem is directly proportional to $N$.
The results show that all devices have comparable performance for the problems considered, 
at least according to the $P_\text{sym}$ metric.
A comparison of the results with and without local fields for both DWC and DWC [NASA] shows that the inclusion of the fields 
slightly increases $P_\text{sym}$, particularly for large $N$.
This is as expected since local fields act as biases to the spins and make the problems easier to solve.
%A comparison of the blue and red curves shows that 
%$P_\text{sym}$ for the DWC is marginally higher than that for the DWP,
%with the difference becoming more pronounced as $N$ increases.
%This suggests a slight increase in performance over the DWP.

FIG.~\ref{fig:success_prob_2} compares $P_\text{sym}$ for the three current-generation devices for $16 \times N$ graphs.
Here, the couplings were drawn from the Sidon set $S_{28}$ while the local fields were set to zero.
The results do not show any noticeable deviations outside the error bars.
Although one would expect the lower-noise device to perform better,
we do not observe an improved performance, at least for the problem class considered.
 
\begin{figure}[h] 
  \includegraphics[width=\linewidth]{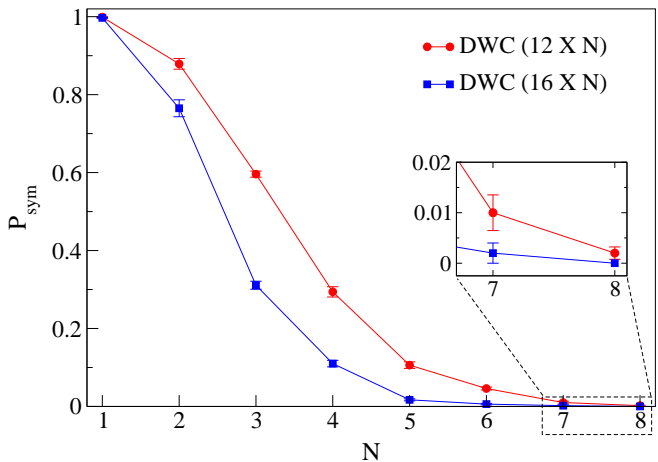}
  \caption{
		Comparison of $P_\text{sym}$ for $12 \times N$ and $16 \times N$ problem graphs on DWC,
		with the number of columns of unit cells $N$ varied from $1$ to $8$.
		The inset shows a magnified view of the data points for $N = 7$ and $N = 8$. 
		The couplings were drawn from the Sidon set $S_{28}$ while the local fields were set to zero.
		For all calculations, the default annealing settings were used.
	  }		
 \label{fig:success_prob_3}
\end{figure}

FIG.~\ref{fig:success_prob_3} shows $P_\text{sym}$ for $12 \times N$ and $16 \times N$ graph sizes on DWC.
$P_\text{sym}$ for $16 \times N$ graphs decreases with increasing $N$ more rapidly than that for $12 \times N$ graphs,
as a result of the increased number of qubits in $16 \times N$ graphs. 
For the largest possible graph size $16 \times 8$, none of the lowest-energy solutions 
out of the $1000$ problem instances satisfied the symmetry requirements, resulting in a $P_\text{sym}$ value of zero.

\begin{figure}[h] 
  \includegraphics[width=\linewidth]{success_prob_2.eps}
  \caption{
		Comparison of $P_\text{sym}$ for $16 \times N$ problem graphs on DWC,
		DWC [NASA], and DWC [lower noise], with the number of columns of unit cells $N$ varied from $1$ to $8$.
		The couplings were drawn from the Sidon set $S_{28}$ while the local fields were set to zero.
		For all calculations, the default annealing settings were used.
	  }		
 \label{fig:success_prob_2}
\end{figure}

During the annealing process, the mirror couplings introduce correlations between the quantum spin states occupying the graph $G$ and its mirror counterpart $G^\prime$. 
These quantum correlations lead to spatial correlations between the solutions returned by the device for $G$ and $G^\prime$. 
When the device fails to return the ground state, 
these spatial correlations can be used as a measure of how close the solutions are to the true ground state.
To demonstrate this, we chose $1000$ problem instances for which none of the lowest-energy solutions out of $1000$ annealing runs 
satisfied reflection symmetry about the mirror plane.
For each lowest-energy solution of each such problem instance,  
we calculated the Hamming distance between each column of unit cells on graph $G$ and the corresponding mirror column on $G\prime$. 
The results for all $1000$ graph instances were averaged to reduce statistical fluctuations. 
These average Hamming distance measurements were further normalized by dividing by the number of functional qubits in the corresponding column of unit cells. 
Such a normalized measurement will yield a value of $0.5$ if the particular column and its mirror counterpart are completely uncorrelated, 
$0$ if the two columns are identical, and $1$ if the two columns satisfy spin-flip symmetry. 

\begin{figure}[h] 
  \includegraphics[width=\linewidth]{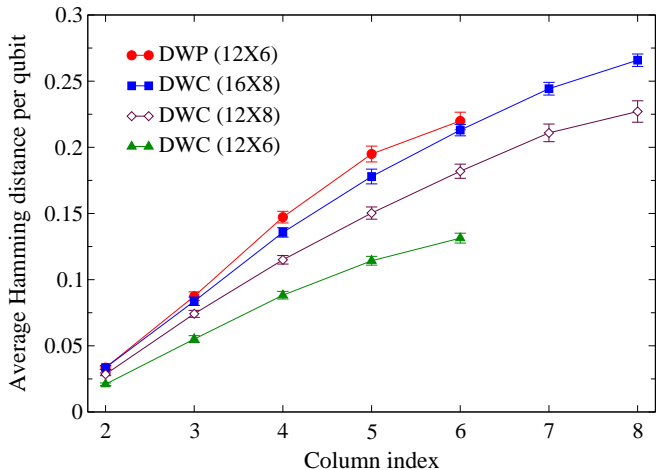}
  \caption{
	  Average Hamming distance between each column of unit cell on graph $G$ and the corresponding mirror column on $G^\prime$, as a function of the column index. 
	  The column indices are counted from the mirror plane and increase with the distance from the mirror plane.
	  For all problem graphs, the couplings were drawn from the Sidon set $S_{28}$ while the local fields were set to zero.
	  The curves are for different graph sizes on DWP and DWC, under default annealing settings.
	  }		
 \label{fig:ham_distance_1}
\end{figure}

FIG.~\ref{fig:ham_distance_1} shows the average Hamming distance as a function of the column index 
for different problem graph sizes on the DWP and DWC devices.
The column indices are counted from the mirror plane and increase with the distance from the mirror plane.
${Index = 1}$ (not shown in the graph) represents the column closest to the mirror plane
which is directly coupled to its mirror counterpart via mirror couplings.
As the spatial correlations get weaker with increasing distance from the mirror plane,
the Hamming distance gradually increases with increasing column index for all curves.
The curves for three different graph sizes ($16 \times 8$, $12 \times 8$, and $12 \times 6$) on the DWC
show that this gradual increase in the Hamming distance successively becomes more pronounced as the problem size increases.
A comparison of the two curves for the DWC and DWP for the $12 \times 6$ graph size clearly shows that
the Hamming distance for the DWP increases more rapidly than that for the DWC.
This leads to an interesting conjecture with regard to the performance of the two devices. 
According to the results shown in FIG.~\ref{fig:success_prob_1}, we do not observe a noticeable
difference in the performance of the DWC and DWP devices in terms of determining the true ground states.
However, the comparison of Hamming distance curves indicates that the lowest-energy solutions provided by the DWC are 
considerably closer to the ground states than the ones provided by DWP.
This suggests that the DWC may indeed outperform the DWP in terms of providing ``near'' optimal solutions,
if not the optimal solution. 

\begin{figure}[h] 
  \includegraphics[width=\linewidth]{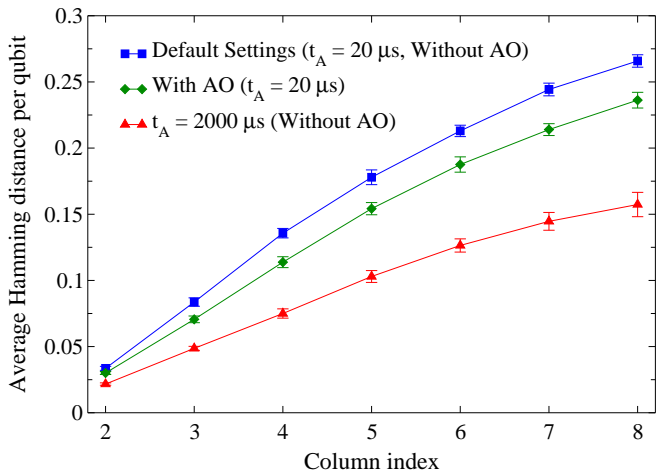}
  \caption{
	  The effect of annealing time ($t_\text{A}$) and annealing offsets (AO) on the the column-wise average Hamming distance for $16 \times 8$ graphs on the DWC.
	  The column indices are counted from the mirror plane and increase with the distance from the mirror plane.
	  For all graphs, the couplings were drawn from the Sidon set $S_{28}$ while the local fields were set to zero.
	  }		
 \label{fig:ham_distance_2}
\end{figure}

D-Wave devices allow the users to make limited adjustments to the default annealing schedule.
One such adjustable parameter is the annealing time ($t_\text{A}$), which can be varied in the range $20-2000$\;$\mu$s on 
both D-Wave 2X and D-Wave 2000Q devices.
The D-Wave 2000Q devices also provide the capability to ``offset'' the annealing paths of individual qubits 
such that the annealing process of certain qubits are delayed/expedited.
Here we investigate how these adjustable parameters affect the column-wise average Hamming distance of the lowest-energy solutions
(See FIG.~\ref{fig:ham_distance_2}).
The blue and red curves respectively show the Hamming distances for $t_\text{A} = 20$\;$\mu$s (default value)
and $t_\text{A} = 2000$\;$\mu$s on the DWC.
The Hamming distance for $t_\text{A} = 20$\;$\mu$s increases more rapidly with the column index than that for $t_\text{A} = 2000$\;$\mu$s,
indicating that the performance of the device increases with increasing annealing time as expected.
The green curve shows the Hamming distance for the DWC with the annealing paths of all the qubits on the left side
of the mirror plane delayed by a normalized offset value of $-0.0866969$.
(Note that the allowable range of offset values differs from qubit to qubit,
and the chosen offset value gives the maximum possible difference in offsets between the qubits on the left and right sides of the mirror plane.)
A comparison with the results obtained for the default settings indicates that
the Hamming distance increases less rapidly with the column index
when the annealing offsets are introduced, suggesting an increase in performance.

\begin{figure}[h] 
  \includegraphics[width=\linewidth]{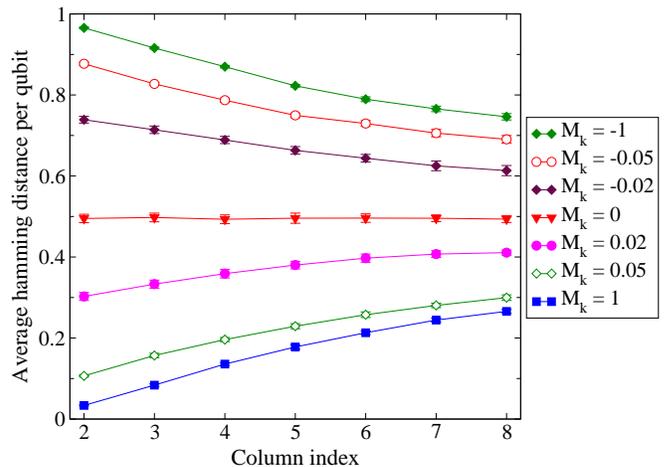}
  \caption{
	  The column-wise average Hamming distance for $16 \times 8$ graphs on the DWC for varying mirror coupling strengths $M_k$.
	  The column indices are counted from the mirror plane and increase with the distance from the mirror plane.
	  For all graphs, the couplings were drawn from the Sidon set $S_{28}$ while the local fields were set to zero.
	  The default annealing settings of the device were used.
	  }		
 \label{fig:ham_distance_3}
\end{figure}

Thus far, our calculations were performed with the mirror couplings $M_k$ set to the maximum possible ferromagnetic value of $+1$.
The magnitude of $M_k$ determines the strength of the spatial correlations between the classical spin configurations 
occupying the graphs $G$ and $G^\prime$. 
To examine the effect of the mirror coupling strength on the spatial correlations,
we calculated the column-wise Hamming distance of the lowest-energy solutions on the DWC for varying values of $M_k$ 
in the range $-1 \leq M_k \leq +1$ (See FIG.~\ref{fig:ham_distance_3}).
In the absence of mirror couplings ($M_k = 0$), the average Hamming distance remains $0.5$ within the error bars, 
indicating that the solutions for $G$ and $G^\prime$ are uncorrelated. 
In the case of the lowest possible antiferromagnetic value $M_k = -1$, the Hamming distance is close to $1$ 
for the column closest to the mirror plane, 
but gradually decreases with the increasing column index as the spatial correlations get weaker.
As $M_k$ is gradually increased from $-1$ to $+1$, we observe a systematic shift of the Hamming distance curves
in accordance with the sign and the magnitude of the corresponding $M_k$ values.
It is interesting to observe the difference from the $M_k<0$ and $M_k>0$ reflection about the Hamming distance 0.5 ($M_k=0$).  
The lack of this expected symmetry may reflect some bias in the D-Wave chip.

\section{Summary}

We have demonstrated a generic approach for benchmarking quantum annealing devices based on symmetry properties associated with 
graphs with mirror symmetry.
This method does not require the prior knowledge of the true ground states of the benchmark problem instances, 
and hence is a more efficient alternative to the conventional benchmarking schemes that rely on direct comparison 
of the solutions to their predetermined ground states. 
In addition, examining Hamming distances of returned states as a function of distance from a mirror plane allows subtle investigations of differences 
in device models and of adjustable parameters on devices.  
In particular, we found for the previous- and current-generation D-Wave models, 
although they do not differ substantially in the probability of obtaining ground states with the required symmetry, 
while the Hamming distance analysis suggests that the current model outperforms the previous model on identical graphs.

Although we have only used mirror symmetry and mirror+spin-flip symmetry on the two ge D-Wave machines, 
our method is a generic approach easily generalized to other situations.  
One generalization would be to quantum annealers with native graph structures that are different from the $K_{4,4}$ Chimera graph of the current D-Wave machines.  
Another generalization would be to use different or additional symmetries, or more than two copies of the graph $G$.  
The method is also easily extended to hypergraphs, wherein more than two-body interactions are present between qubits.
Moreover, the method can be used for arbitrary 2-local Hamiltonians,
which have been shown to belong to QMA-complete complexity class~\cite{two_local}.

As discussed in our earlier preliminary study~\cite{perera:16}, 
our method can also be utilized as an answer checking method for assessing the validity of candidate ground state solutions.
One additional big advantage of the method is for specific applications on quantum annealing machines, 
where for a given algorithm for a specific problem the quantum annealer is used as a part of a classical computation, 
with the quantum annealer solely used to return a state which is hopefully the ground state. 
One such application would be the use of adiabatic quantum computation in quantum chemistry calculations~\cite{quantum_chemistry}.
For example, when using D-Wave machines, 
one usually does not know whether the quantum annealer part of the calculation is a set of QUBO problems which have ground states 
that are easier or more difficult for the (imperfect) quantum annealer to solve.  
Repeated calls to the quantum annealer can be used to overcome this difficulty.  
The measurement of $P_\text{sym}$ allows one to measure the difficulty of the class of QUBO problems for the specific application.  
Furthermore, by throwing away any solution which does not have the required symmetry, 
one can be much more confident that the QUBO solutions with the symmetry is a ground state, 
thereby increasing the usefulness of the quantum annealer as the mechanism to find the ground state solution to the QUBO.
This type of error-correcting using the symmetry therefore will make the entire quantum+classical computation work more efficiently.  
If $P_\text{sym}$ is too small or zero, there is insufficient error correcting ability on the imperfect quantum annealer to solve the particular problem.
In contrast, a sufficiently large $P_\text{sym}$ value would suggest that the quantum annealer is capable of providing solutions that are close to the true
ground state of the QUBO problem,
and consequently the entire quantum+classical calculation is sufficient to solve the problem.

\begin{acknowledgments}
This research was sponsored by Pacific Northwest National Laboratory (PNNL). 
MAN acknowledges partial support from a Fulbright Distinguished Chair grant from the Czech J.W. Fulbright Commission. 
Time on the D-Wave 2X was granted through Universities Space Research Association (USRA).
We sincerely thank D-Wave Systems Inc. for providing computing time on their latest D-Wave 2000Q machine. 

\end{acknowledgments}

\bibliography{dwave.bib}

\end{document}